\documentclass[applsci,article,submit,pdftex,moreauthors]{Definitions/mdpi} 

\firstpage{1} 
\pubvolume{1}
\issuenum{1}
\articlenumber{0}
\pubyear{2024}
\copyrightyear{2024}
\datereceived{ } 
\daterevised{ } 
\dateaccepted{ } 
\datepublished{ } 
\hreflink{https://doi.org/} 

\usepackage{graphicx}
\usepackage{amsmath}
\usepackage{amsthm}
\usepackage{tabularx}
\usepackage{tabularray}
\usepackage{rotating}
\usepackage{musixtex}
\microtypesetup{protrusion=false}
\preto{\abstractkeywords}{\nolinenumbers}

\Title{DExter: Learning and Controlling Performance Expression with Diffusion Models}

\TitleCitation{Title}

\Author{Huan Zhang $^{1}$\orcidA{}, Shreyan Chowdhury $^{2}$\orcidC{}, Carlos Eduardo Cancino-Chacón $^{2}$\orcidF{}, Jinhua Liang$^{1}$\orcidE{}, Simon Dixon $^{1}$\orcidB{}, Gerhard Widmer$^{2}$\orcidD{}}

\AuthorNames{Huan Zhang, Shreyan Chowdhury, Carlos Eduardo Cancino-Chacón, Jinhua Liang, Simon Dixon and Gerhard Widmer}

\AuthorCitation{Zhang, H.; Chowdhury, S.; Cancino-Chacón, C. E.; Liang, J.; Dixon, S; Widmer, G}

\address{%
$^{1}$ \quad Centre for Digital Music, Queen Mary University of London\\
$^{2}$ \quad Institute of Computational Perception, Johannes Kepler Universität Linz}

\corres{Correspondence: huan.zhang@qmul.ac.uk}

\abstract{In the pursuit of developing expressive music performance models using artificial intelligence, this paper introduces DExter, a new approach leveraging diffusion probabilistic models to render Western classical piano performances. In this approach, performance parameters are represented in a continuous expression space and a diffusion model is trained to predict these continuous parameters while being conditioned on the musical score. Furthermore, DExter also enables the generation of interpretations (expressive variations of a performance) guided by perceptually meaningful features by conditioning jointly on score and perceptual feature representations. Consequently, we find that our model is useful for learning expressive performance, generating perceptually steered performances, and transferring performance styles. We assess the model through quantitative and qualitative analyses, focusing on specific performance metrics regarding dimensions like asynchrony and articulation, as well as through listening tests comparing generated performances with different human interpretations. Results show that DExter is able to capture the time-varying correlation of the expressive parameters, and compares well to existing rendering models in subjectively evaluated ratings. The perceptual-feature-conditioned generation and transferring capabilities of DExter are verified by a proxy model predicting perceptual characteristics of differently steered performances.}

\keyword{Piano Performance, Expressive Rendering, Diffusion Model, Deep Learning, Music} 

\begin{document}

\section{Introduction}

A trained musician can take a piece of music and interpret it in their own way, moulding and varying the emotional expression of the piece by subtly changing performance parameters. 
Parametric dimensions include timing, dynamics, articulation, and use of devices like sustain pedals in piano. Studying such expression patterns has long been of keen interest to musicians, educators and researchers, and it presents a compelling inquiry into exploring whether such intricate expressions can be accurately encapsulated and replicated by computational systems \cite{Widmer2003InFactor}. The accurate replication of human musical expression by machines not only bridges the gap between traditional and technological approaches to music but also opens new avenues for interactive performances \cite{Cancino-Chacon2023ACCompanionAccompanist} and music education systems \cite{Morsi2024SimulatingContexts}. Leveraging such technology can enhance musical training, allowing students and professionals alike to experiment with and learn from dynamically generated expression, thus broadening both creative perspectives and educational methods.  
In this paper, we aim at rendering such an expressive performance of a piece of music from its score using a machine learning model. We propose \textbf{DExter}, a \textit{D}iffusion-based \textit{Ex}pressive performance genera\textit{t(o)r}, which predicts expression parameters conditioned on the score. In addition, we investigate whether the rendering process can also be conditioned on desired high-level performance characteristics given in the form of \textit{mid-level perceptual features} \cite{Aljanaki2018Data-drivenModeling,Chowdhury2019}, in this way permitting us to control general performance character, as well as to explore the potential of style transfer within the varied space of human expressive performances.
In this context, we leverage the conditional design of diffusion models as well as the diffusion chain to serve as a mechanism to regulate the extent of transferred information from a source performance to a target performance.


This paper offers three contributions\footnote{Project demo page with examples: \url{http://bit.ly/4a1xs1x}. Code is available at \url{https://github.com/anusfoil/DExter}.}: 1) we propose a diffusion-based method for learning and conditioning the expression parameters in Western classical solo piano performance; 2) we conduct a comprehensive quantitative evaluation on the rendered outputs along with other renderers in the literature (re-trained to make for a fair comparison), taking into account multiple expressive dimensions such as asynchrony and articulation; and 3) we explore mid-level conditioned generation and style transfer with our model and conduct an experimental study on the conditioning effects.

\section{Related Work}
\subsection{Expressive Performance Rendering}

Expressive performance rendering has long been a challenge for Music Information Retrieval (MIR) research. While the role of machine learning in such a task was recognised early on \cite{Widmer2003InFactor}, several rule-based methods have been proposed and investigated over the years \cite{widmer2004computational,Cancino-Chacon2018ComputationalReview,Kirke2013GuidePerformance}. 
Early experiments in deep-learning based performance rendering \cite{CancinoChacon2018ComputationalModels,Maezawa2019RenderingRNN,Jeong2019VirtuosoNetPerformance} use traditional sequence modeling architectures like RNNs and LSTMs with modifications focusing on the music hierarchy and score features being applied as inputs. 
Recently, transformer-based systems \cite{Rhyu2022SketchingLearning,Borovik2023ScorePerformerControl} have been proposed for controllable rendering, predicting different aspects of performance such as the shape of expressive attributes \cite{Rhyu2022SketchingLearning} and performance direction markings in the score \cite{Borovik2023ScorePerformerControl}. All of the above systems predict descriptors designed to capture expressive aspects of musical performance, typically tokens representing local tempo and timing deviations. However, such tokenized and quantized encodings of performance parameters are not lossless and can result in a large vocabulary to train \cite{Zhang2023DisentanglingPerformance}. 

Regarding the \textit{evaluation} of performance rendering systems, there has been a growing criticism of the practice of evaluating against a single ground truth and ignoring the variations in interpretations \cite{Peter2023SoundingPerformance}, as reconstruction-based error analysis has inherent limitations on fidelity and diversity \cite{Plasser2023DiscreteGeneration,Peter2023SoundingPerformance}. 
To mitigate this problem, we will evaluate the rendered performances with respect to a multitude of performance parameter dimensions, and against multiple different human interpretations.

\subsection{Diffusion Models in Music}

Diffusion Probabilistic Models (DPMs) generate data by inverting a Markovian data corruption process. 
DPMs have demonstrated impressive results first in the vision domain by generating text-controlled images \cite{Ramesh2021Zero-ShotGeneration}, and then also in the audio domain, with the most promising applications involving generation of high-fidelity audio samples \cite{Kong2021DiffwaveSynthesis,Chen2021WavegradGeneration} and synthesis of speech \cite{Kim2022Guided-TTSGuidance} and music \cite{Hawthorne2022Multi-instrumentDiffusion}. 

Symbolic music, however, seems to be a more challenging target for DPMs -- the challenge is to fit their probabilistic formulation into discrete data distributions. \citet{Mittal2021SymbolicModels} train continuous DDPMs (Denoising Diffusion Probabilistic Models) on sequences of latent MusicVAE \cite{Roberts2018HierarchicalMusic} embeddings, in order to achieve generation of monophonic melodies. \citet{Plasser2023DiscreteGeneration} build upon the MusicVAE-like token representation and directly apply discrete denoising diffusion probabilistic models (D3PM). Another representation suitable for learning symbolic music \cite{Zhang2023SymbolicEvaluation} under the DPM framework is the piano roll: \citet{Cheuk2023DiffrollCapability} managed to transcribe music by generating a piano roll using an audio spectrogram as condition. \citet{Min2023PolyffusionControls} also achieved piano roll generation with more diverse control such as infilling music context and high-level guidance of chords and texture. 


Our work places music DPMs into a niche spot: while the rendering is applied on symbolic data (discrete notes), DExter predicts continuously varying expressive parameters which are then applied at the note level. Generation of the continuous expressive parameters facilitates fine-grained control of performance parameters of each note without the reverse diffusion process having to learn a quantized representation space.

\begin{figure*}
    \centering
    \includegraphics[width=\linewidth]{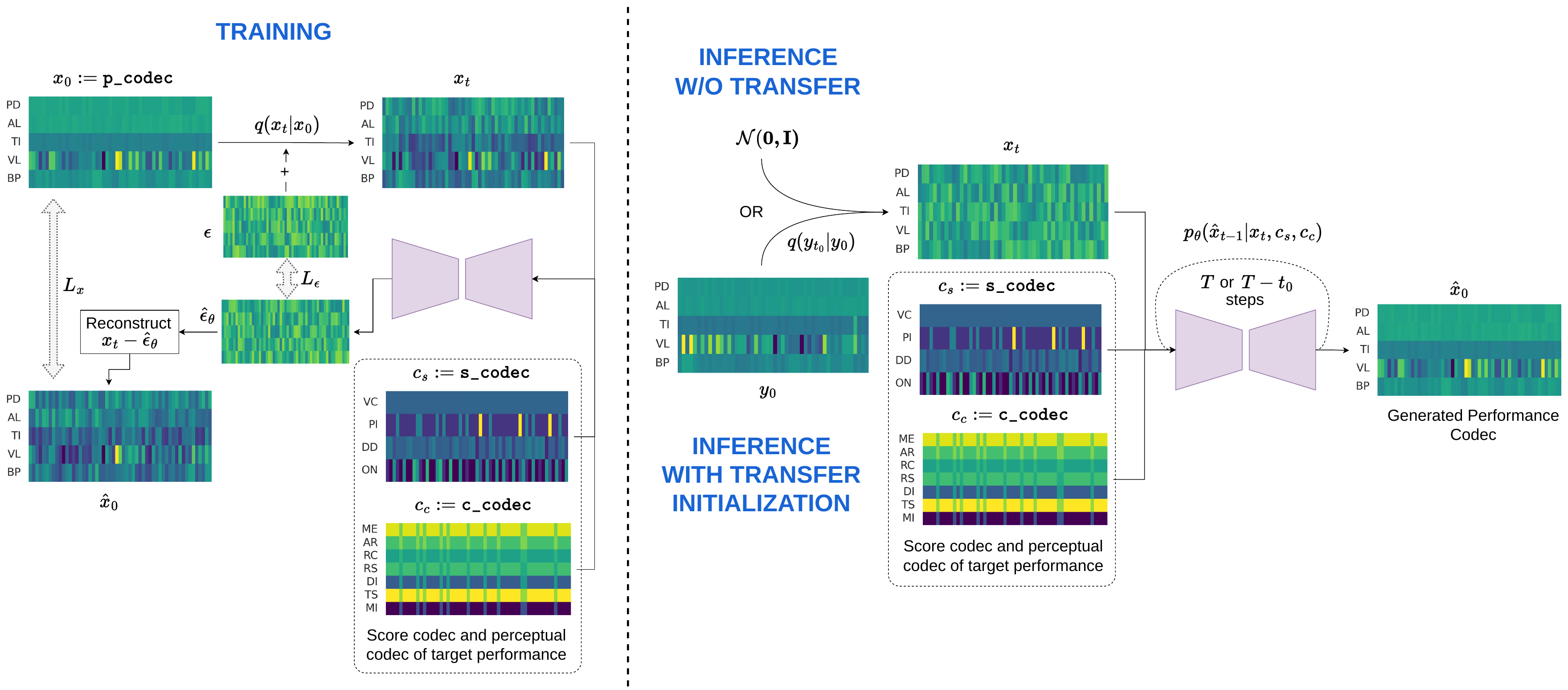}
    \caption{Training (left) and inference (right) phases of the diffusion framework. Training starts with \texttt{p\_codec} and corrupts the $x_0$ by injecting noise; the UNet model takes in the corrupted $x_t$, conditions $c_s$ and $c_c$ to predict the injected noise, which is then used to reconstruct $\hat{x}_0$. Loss is calculated for both noise prediction and \texttt{p\_codec} reconstruction. The inference process (right) starts with a random sample from $\mathcal{N}(\mathbf{0}, \mathbf{I})$; the model iteratively predicts the noise and reconstructs $\hat{x}_0$, conditioned on the same $c_s$ and $c_c$. Alternatively for transferal, we initialize the process from another performance $y_0$, corrupting it for $t_0$ steps and denoising for the remaining $T-t_0$ steps.
    }
    \label{fig:diffusion_diagram}
\end{figure*}

\section{Methodology}
\label{sec:methodology}
In this section, we first introduce the representations for expressive parameters, musical score, and perceptual features used to train DExter. We call these \textit{codecs}, which are described in detail in Section~\ref{sec:codecs}. We then explain our diffusion framework that learns these representations, followed by the training and inference architectures and conditioning methodology.

\subsection{The Codecs}
\label{sec:codec_intro}
We represent score information (note onset, duration, pitch, and voice), performance parameters (beat period, velocity, timing, articulation ratio, and pedal), and mid-level perceptual features (melodiousness, articulation, rhythmic complexity, rhythmic stability, dissonance, tonal stability, and minorness) as two-dimensional spectrogram-like matrices of (mostly real-valued, except for the score codec) numeric values. We call these the \textbf{score codec} (\texttt{s\_codec}), the \textbf{performance codec} (\texttt{p\_codec}), and the \textbf{perceptual features codec} (\texttt{c\_codec}) respectively. The task of our diffusion model is to predict a \texttt{p\_codec} conditioned on the \texttt{s\_codec} and \texttt{c\_codec}. Detailed descriptions of the composition of the codecs are given in Section~\ref{sec:codecs}.

\subsection{Diffusion Framework}
\label{sec:diffusion_framework}

We frame the expression rendering problem as the task of learning a continuous space of performance expression parameters. Diffusion models \cite{Ho2020Denoisingmodels} consist of two processes: i) a forward process that transforms each data sample into a standard Gaussian noisy sample step-by-step with a predefined noise schedule; and ii) a reverse process where the model learns to denoise pure-noise inputs gradually, generating samples from the learned training data distribution. In effect, our model aims to convert Gaussian noise $x_t$ into a posterior performance codec $\hat x_0$, conditioned on a score codec $c_s :=$ \texttt{s\_codec}, and a perceptual features codec $c_c :=$ \texttt{c\_codec}.

The \textbf{diffusion forward pass} $q(x_t|x_0)$ produces a noisy version of the performance codec. With the noise $\epsilon \in  \mathcal{N}(\mathbf{0}, \mathbf{I})$ sampled from a standard Gaussian distribution, we blend it with the input sample $x_0$ using $\beta$ as a scaling factor intended to ultimately achieve zero mean and unit variance of the fully-noised result. Specifically, the sampling process applies a linear noise schedule with $\beta \in [0.0001, 0.2]$. As we would like to perform multiple steps simultaneously, reparameterization is applied to derive a closed-form equation, given that $\alpha_{t} = 1 - \beta_{t}$ and $\bar\alpha = \prod^{t}_{s=1}\alpha_{s}$.
\begin{equation}
\label{eq:q_sampling}
    x_t = \sqrt{\bar \alpha_{t}}x_0 + \sqrt{1 - \bar \alpha_{t}}\epsilon
\end{equation}

During \textbf{training}, model $f_{\theta}(x_t, t, c_s, c_c)$ learns to predict the injected noise $\hat \epsilon_{\theta}$ given a random timestep $t$ and its noised codec version $x_t$ calculated in the forward pass. $t$ is sampled from $[1,T]$; we use $T=1000$ in our experiments. Then, we use the predicted noise $\hat \epsilon_{\theta}$ to reconstruct the predicted initial codec $\hat x_0$ by inverting Eq.~\ref{eq:q_sampling}. 

The training objective combines the noise estimation and codec reconstruction as shown in Eq.~\ref{eq:loss}. Although the noise prediction is theoretically enough for training the model, empirically we found that constraining on the reconstructed codec yields better performance, with weighting $h=0.2$. 

\begin{equation}
\label{eq:loss}
    L(\theta) = \parallel \epsilon - \hat\epsilon_{\theta} \parallel^2 + h \parallel x_0 - \hat x_0 \parallel^2
\end{equation}

During \textbf{inference}, we start from a Gaussian noise distribution $p(x_t) \sim  \mathcal{N}(\mathbf{0}, \mathbf{I})$ and iteratively generate the codec posterior through  $p_{\theta}(\hat x_{t-1}|x_{t}, c_s, c_c) = \mathcal{N}(\mu_{\theta, t}(x_t, t, c_s, c_c), \sigma^2_{t}\mathbf{I})$ until $\hat x_0$ is reached. As the model $f_{\theta}$ estimates noise $\hat \epsilon_{\theta}$, we use it to construct the model mean $\mu_{\theta, t}$ and the posterior variance is predetermined by the noise schedule. Full construction of model mean and posterior variance is given in Eq.~\ref{eq:model_mean} and Eq.~\ref{eq:posterior_variance}:
\begin{equation}
\label{eq:model_mean}
    \mu_{\theta, t}(x_t, t, c_{s}, c_{c}) = \sqrt{\frac{1}{\alpha_t}} (x_t - \frac{\hat\epsilon_{\theta}(1 - \alpha_t)}{\sqrt{1 - \bar\alpha_t}}) 
\end{equation}
\begin{equation}
\label{eq:posterior_variance}
    \sigma^2_{t} = (1 - \alpha_t)\frac{1 - \bar\alpha_{t-1}}{1 - \bar\alpha_{t}}
\end{equation}

\begin{figure}[t]
    \centering
    \includegraphics[trim={0 2cm 0 2cm}, clip, width=\linewidth]{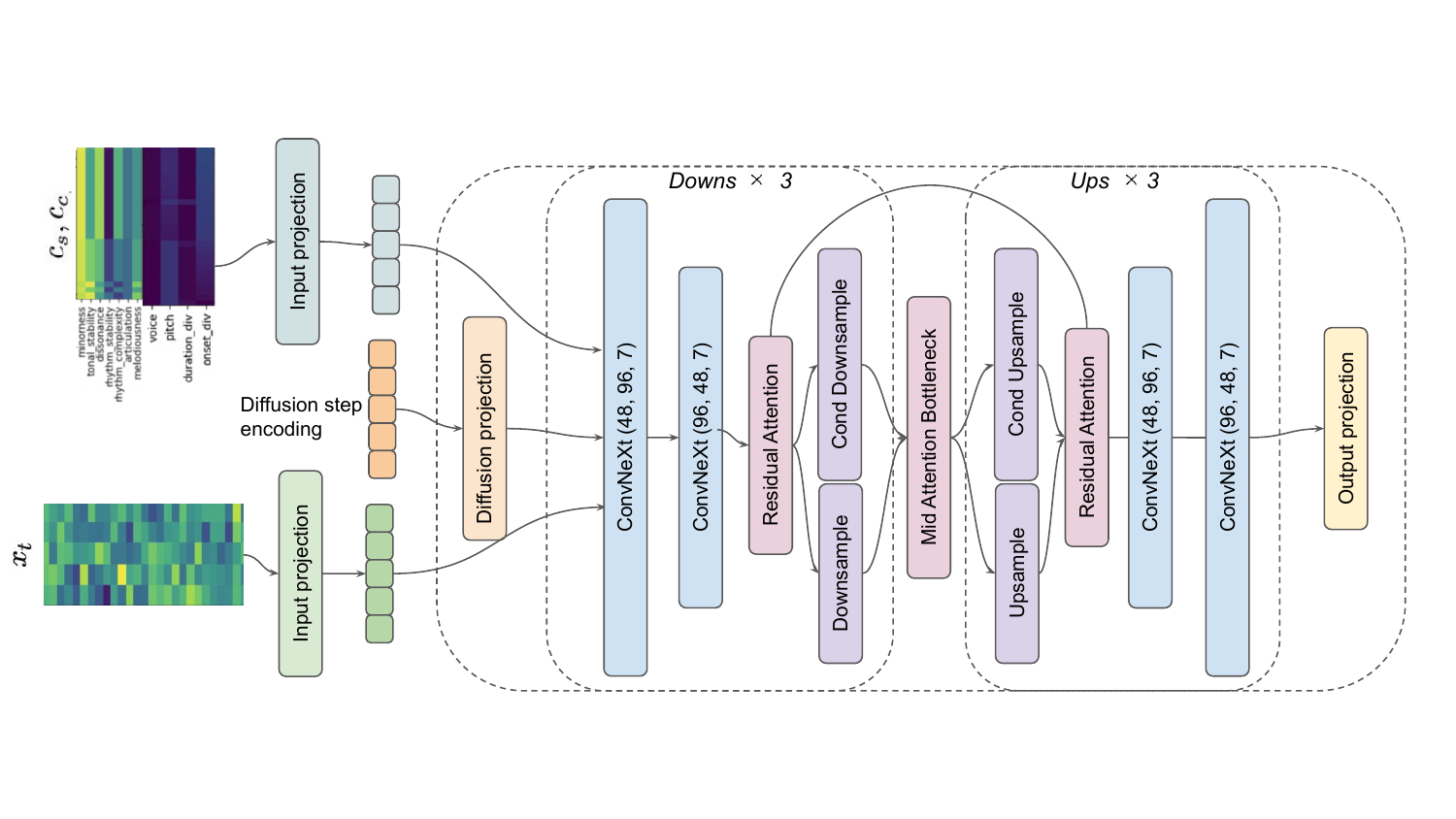}
    \caption{Diagram of the UNet conditioning module in the network.}
    \label{fig:layer_structure}
\end{figure}

\subsection{Architecture and Conditioning}
\label{sec:conditioning}

We employ a 2-D UNet as the backbone of our architecture; the detailed layer and insertion structure can be found in Fig.~\ref{fig:layer_structure}. The conditioning on score information and perceptual feature information is enforced by a joint conditioning layer that projects the score dimensions and perceptual dimensions (5 and 7 respectively -- see definition of codecs below) onto 512 dimensions. The diffusion timestep $t$ is encoded via sinusoidal position embeddings. The input codec and the conditioning codecs are downsampled and upsampled through ResNet blocks and 2D convolutions. Attention layers are interleaved at bottlenecks.

Before narrowing down on the above described architecture, we experimented with a DiffWave-based architecture \cite{Kong2021DiffwaveSynthesis} which uses a series of 12 residual layers of 1D convolution. For conditioning, we also experimented with FiLM \cite{Perez2018FiLMlayer,Kim2023DiffVelModel} which yields comparable results to the UNet model. We found that our present architecture gives the best trade-off between model simplicity and performance. 

\setlength{\abovedisplayskip}{1pt}
\setlength{\belowdisplayskip}{1pt}

\textbf{Classifier-Free Guidance (CFG)} \cite{Ho2022Classifier-FreeGuidance} is widely used for conditioning diffusion models to achieve controllable generation, which we also adopt. During training, a dropout layer is applied to the conditions $c_s$ and $c_c$ to randomly mask out the conditions with probability $p$, in order to simultaneously train the conditional model $f_{\theta}(x_t, t, c_{s}, c_{c})$ and unconditional model $f_{\theta}(x_t, t)$.  We fixed $p=0.1$ in our training. In inference, a weight parameter $w$ is applied as the guidance scale to a combined prediction.
\begin{equation}
\label{eq:weight}
    \hat \epsilon = w\hat\epsilon(x_t, t) + (1-w)\hat\epsilon(x_t, t, c_s, c_c)
\end{equation}



\begin{table*}
    \scriptsize
    \centering
    \begin{tabularx}{\linewidth}{p{2cm} c c c c X}
        \toprule
        Dataset & Pieces & Performances & Duration & MIDI & Repertoire \\ [0.5ex]
        \midrule
        Vienna4x22 (\cite{vienna4x22}) & 4 & 88 & 2h 18m & recorded & Excerpts from 4 pieces by F. Chopin (Op.~10 No.~3, Op.~38), W. A. Mozart (KV331, first mov.), and F. Schubert (D.~783 No.~15) \\
        \addlinespace 
        (n)ASAP (\cite{Peter2023AutomaticDataset})  & 235 & 1067 & 94h 30m  & recorded & Common Practice Period solo piano works by 15 composers \\
        \addlinespace 
        ATEPP-subset* (\cite{Zhang2022ATEPPPerformance}) & 1580 & 11677 & 1000h & transcribed  & Solo piano works by 25 composers, ranging from Baroque to Modern era \\
        \bottomrule
    
    \end{tabularx}
    \caption{Overview of datasets used in experiments.}
    \label{tab:datasets_overview}
\end{table*}

\section{Data, Representation, and Processing}
\label{sec:data}

\subsection{Input and Target Encodings}
\label{sec:codecs}

The \textbf{performance codec} (\texttt{p\_codec}) -- our prediction target -- was originally proposed in the expressive rendering framework \textit{Basis Mixer} \cite{CancinoChacon2018ComputationalModels}, where four expressive parameters are computed for each note $n$ appearing in the score.
These parameters of the \texttt{p\_codec} encoding the expression controls modify properties of notes in a MIDI piano performance, thus changing speed and loudness of the performance with time. Combined with score information, the full expressive performance can be reconstructed in a lossless fashion. We expanded the original Performance Codec v.1.0 \cite{CancinoChacon2018ComputationalModels,Cancino-Chacon2023ACCompanionAccompanist} by defining an additional parameter for sustain pedal control. The resulting five (note-wise) performance parameters are as follows:
\begin{itemize}
    \item Beat period: the ratio of the inter-onset intervals (IOI) between two consecutive notes of the performance and the score. This parameter is
    computed for each onset $o_k$ instead of each note $n_i$. It is defined as: \\$x_\mathit{bp}(o_k) = \frac{\text{IOI}_{\mathit{perf}}(o_k)}{\text{IOI}_{\mathit{score}}(o_k)} = \frac{\hat o^{\mathit{perf}}_{k+1} - \hat o^{\mathit{perf}}_{k}}{o_{k+1} - o_k}$, where $\hat o^{\mathit{perf}}_{k}$ is the actual performed onset time, in seconds, corresponding to score onset $o_k$ in beats,
    calculated as the average onset time of all notes played at score onset position $o_k$.  
    \item Velocity: $x_\mathit{vel}(n_i) = \frac{\text{vel}(n_i)}{127}$, where vel is the MIDI velocity of a played note.
    \item Timing: $x_{\mathit{tim}}(n_i) = \Delta_t(n_i) = \hat o^{\mathit{perf}}(n_i) - \text{onset}(n_i)$, the average onset time of all notes played at score onset position (used in beat period) minus the performance onset time of $n_i$. Taking beat period as the `tempo grid' notion \cite{Gillick2021DrumrollGrids}, timing would then refer to the micro-deviation of each note relative to the grid. 
    \item Articulation ratio: $x_{\mathit{art}}(n_i) = \frac{\text{dur}^{\mathit{perf}}(n_i)}{\text{dur}(n_i)\cdot x_\mathit{bp}(n_i)}$, measures the fraction of the expected note duration that is actually played.
    \item Pedal: $x_\mathit{ped}(n_i) = \frac{\text{ped}(n_i)}{127}$, where $\text{ped}(n_i)$ is the discrete MIDI pedal value at the onset of $n_i$. Note that pedal encoding is not lossless since  changes of value between note onsets will not be captured.
\end{itemize}

The \texttt{p\_codec} is fully invertible in that the full event information from the MIDI file [\texttt{Pitch}, \texttt{Onset}, \texttt{Duration}, \texttt{Velocity}] can be reconstructed given the \texttt{p\_codec} and score \texttt{s\_codec}.

The \textbf{Score codec} (\texttt{s\_codec}) represents the musical score and is derived from the note array from the \textit{partitura} package \cite{Cancino-Chacon2022PartituraProcessing}. Aligned with the \texttt{p\_codec} at the note level, it contains four score parameters for each note: (notated) Onset, Duration, Pitch, and Voice, resulting in a 2D matrix of dimension $4\times n$ where n is the number of notes. The score is indispensable for performance conditioning, as it defines the musical content of the piece.



The \textbf{Perceptual features codec} (\texttt{c\_codec}), which we use as steering inputs for the performance generation, are representations of the so-called \textit{mid-level perceptual features} \cite{Aljanaki2018Data-drivenModeling}, namely -- \textit{melodiousness}, \textit{articulation}, \textit{rhythm complexity}, \textit{rhythm stability}, \textit{dissonance}, \textit{tonal stability}, and \textit{`minorness}' (or \textit{mode}). They describe musical qualities that most listeners can easily perceive. Taking cue from previous research \cite{Chowdhury2019,Chowdhury2022Thesis} showing that these features effectively represent musical factors underlying a wide range of emotions and capture variations in expressive character between different performances of a piece \cite{Chowdhury2021OnFeatures}, we incorporate these as the performance steering inputs. In our scenario, these features are calculated by a previously trained specialised model \cite{Chowdhury2019}, over the recorded audio performance data of  Vienna$4\times22$, (n)ASAP, and ATEPP datasets (see Sec.~\ref{sec:datasets_used}).  
The values are calculated from successive overlapping 15s windows with hop size of 5s. Each computed window is then aligned with the score note array to broadcast into \texttt{c\_codec}, a 2D matrix of dimension $7\times n$. 



\subsection{Processing}

Given that there could be slight variations in each performance (missing and  extra notes relative to the score), we perform padding based on the score note array so that each pair of performances is perfectly aligned.  To accommodate pieces of different lengths, we train our network on segments of $N$ notes where shorter segments are padded. In our experiment, we take $N=200$ (which corresponds to about 10 to 20 seconds of music depending on the tempo and note density).

\subsection{Mixup Augmentation}
\textit{Mixup} \cite{Zhang2018MixupMinimization} is a data augmentation scheme that regularizes a network to favor simple linear behavior between training examples. To strengthen our model's ability to model different interpretations, we fuse \texttt{p\_codec} pairs $x_1$ and $x_2$ (codecs representing two different performances of the same piece) and their corresponding \texttt{c\_codec} using Eq.~\ref{eq:mixup}, where $\lambda$ is a scaling factor varying between $[0, 1]$. 
\begin{equation}
\label{eq:mixup}
    x_{1, 2} = \lambda x_1 + (1 - \lambda) x_2
\end{equation}
After the mixup augmentation, our dataset consists of 170k segments; the interpolated data are only used in training.

\subsection{Datasets and Training Setup}
\label{sec:datasets_used}
We used three datasets of expressive performances (from the Western classical music solo piano repertoire): Vienna$4\times22$ \cite{vienna4x22}, (n)ASAP \cite{Peter2023AutomaticDataset}, and ATEPP \cite{Zhang2022ATEPPPerformance}. Each dataset includes audio, performance MIDI, score in MusicXML format, and their alignment. Information and a comparison of these sets can be found in Table~\ref{tab:datasets_overview}. The training is based on ATEPP and 80\% of (n)ASAP and Vienna$4\times22$ data, while the testing set (used in all subsequent experiments in Sec.~\ref{sec:evaluation}) contains the remaining 20\% of (n)ASAP and Vienna$4\times22$ data. The latter two datasets were recorded on computer-controlled grand pianos and are thus more accurate and precise than the ATEPP data, which were obtained through curated audio transcription.

For the training of the network as mentioned in Sec.~\ref{sec:methodology}, we use the Adam optimizer with a learning rate $5\times10^{-5}$, and employ early stopping with a patience of 50 epochs.

\renewcommand{\arraystretch}{1.5}

\begin{table*}
\centering
\begin{tabular}{lccc} 
\toprule
& Basis Mixer \cite{CancinoChacon2018ComputationalModels} & VirtuosoNet \cite{Jeong2019VirtuosoNetPerformance} & DExter (Ours) \\
\midrule
\textit{Deviation multiple} ($\rightarrow$0) \\[-1ex]
\midrule
articulation.key\_overlap\_ratio & \textbf{0.62$\pm$3.15} & 1.92$\pm$2.72 & 2.24$\pm$2.76 \\
asynchrony.pitch\_correlation & -1.19$\pm$1.41 & -1.67$\pm$1.25 & \textbf{-1.17$\pm$1.66} \\
asynchrony.delta & \textbf{4.10$\pm$1.78} & 4.38$\pm$1.63 & 4.43$\pm$2.21 \\
dynamics.agreement & -0.07$\pm$1.33 & -0.40$\pm$1.30 & \textbf{-0.002$\pm$1.05} \\
dynamics.consistency & \textbf{-0.67$\pm$1.64} & -1.07$\pm$1.56 & 0.73$\pm$2.21 \\
dynamics.ramp\_correlation & -0.36$\pm$2.54 & 0.65$\pm$1.96 & \textbf{-0.28$\pm$2.44} \\
pedal.onset\_value & - & -1.16$\pm$1.68 & -1.39$\pm$2.13 \\
\textbf{tempo\_curve} & \textbf{-0.10$\pm$2.44} & 0.52$\pm$2.48 & 0.72$\pm$2.65 \\
\textbf{velocity\_curve} & -0.67$\pm$1.48 & \textbf{0.15$\pm$1.02} & 1.48$\pm$2.12 \\
\midrule
\textit{KL Divergence} ($\downarrow$) \\[-1ex]
\midrule
articulation.key\_overlap\_ratio & \textbf{0.92$\pm$2.15} & 1.66$\pm$6.89 & 1.64$\pm$3.63 \\
asynchrony.pitch\_correlation & 0.14$\pm$0.278 & \textbf{0.13$\pm$1.25} & 0.20$\pm$0.33 \\
asynchrony.delta & 4.04$\pm$5.15 & 4.83$\pm$9.58 & \textbf{1.29$\pm$3.16} \\
dynamics.agreement & 0.10$\pm$0.04 & 0.09$\pm$0.04 & \textbf{0.06$\pm$0.04} \\
dynamics.consistency & 0.12$\pm$0.23 & \textbf{0.06$\pm$0.07} & 0.28$\pm$0.49 \\
dynamics.ramp\_correlation & 1.54$\pm$5.43 & \textbf{0.35$\pm$1.01} & 0.42$\pm$1.13 \\
pedal.onset\_value & - & 0.34$\pm$1.45 & 0.33$\pm$0.36 \\
\textbf{tempo\_curve} & 0.98$\pm$2.55 & \textbf{0.65$\pm$1.86} & 1.26$\pm$5.66 \\
\textbf{velocity\_curve} & 0.16$\pm$0.21 & \textbf{0.10$\pm$0.06} & 0.71$\pm$1.37 \\
\midrule
\textit{Pearson's Correlation} ($\uparrow$) \\[-1ex]
\midrule
articulation.key\_overlap\_ratio & -0.01$\pm$0.13 & 0.05$\pm$0.16 & \textbf{0.11$\pm$0.17} \\
asynchrony.pitch\_correlation & 0.33$\pm$0.25 & 0.55$\pm$0.17 & \textbf{0.57$\pm$0.25} \\
asynchrony.delta & 0.17$\pm$0.22 & \textbf{0.29$\pm$0.19} & 0.28$\pm$0.21 \\
dynamics.agreement & 0.02$\pm$0.87 & 0.04$\pm$0.84 & \textbf{0.11$\pm$0.79} \\
dynamics.consistency & 0.92$\pm$0.17 & 0.91$\pm$0.13 & \textbf{0.92$\pm$0.15} \\
dynamics.ramp\_correlation & 0.04$\pm$0.76 & 0.12$\pm$0.80 & \textbf{0.14$\pm$0.73} \\
pedal.onset\_value & - & 0.01$\pm$0.13 & 0.02$\pm$0.14 \\
\textbf{tempo\_curve} & 0.02$\pm$0.13 & 0.09$\pm$0.19 & \textbf{0.19$\pm$0.17} \\
\textbf{velocity\_curve} & 0.08$\pm$0.23 & 0.21$\pm$0.27 & \textbf{0.27$\pm$0.23} \\
\bottomrule
\end{tabular}
\caption{Quantitative expression metrics in the categories of articulation, asynchrony, dynamics, pedaling, plus global tempo and velocity curves. Columns represent different models, and rows are divided into blocks according to the three different evaluation metrics, with each block detailing the outcomes for all features. Note that each generated performance is compared with multiple human ground truth interpretations.}
\label{tab:quant_expression}
\end{table*}

\section{Evaluation}
\label{sec:evaluation}

In this section, we present quantitative evaluation of generated performances without and with steering, followed by evaluation of performance transfer, and an investigation into the effect of varying the conditioning weight. Finally, we also describe our qualitative study employing a listening test and human participants and present the results.


\subsection{Quantitative Evaluation}
\label{subsec:quant_analysis}

In this subsection, we evaluate our generated samples' expressiveness by comparing core expression attributes with ground truth performances. This experiment is conducted on the aforementioned testing set, with condition of \texttt{s\_codec} and audio performance inferred \texttt{c\_codec} as described in Sec.~\ref{sec:codecs}.
With respect to the critique of reconstruction-based evaluation \cite{Peter2023SoundingPerformance}, we compare with various interpretations of ground truth (the testing set consists of about 5.3 human performances for each piece, on average).


\subsubsection{Assessed Attributes}

The expression attributes we assess are derived from the tempo and velocity curves (joint-onset level), joint-onset asynchrony, articulation, dynamics and pedalling. While a detailed documentation of the selected attributes can be found on the project page, we provide a summary below: 
\begin{itemize}
    \item Tempo curve: Onset-level tempo (inverse of local inter-beat-intervals), with values averaged across notes on the joint onset. (\textbf{tem\_curve})
    \item Velocity curve: Onset-level velocity, with values averaged across notes on the joint onset. (\textbf{vel\_curve})
    \item Asynchrony: The absolute difference in seconds between the earliest and latest note in a joint onset (\textbf{asy.delta}). We also measure the pitch correlation (\textbf{asy.p\_cor}) between the pitch and micro-timing within the joint onset, inspired by the \textit{melody lead} phenomenon \cite{Goebl2001MelodyArtifact}.
    \item Articulation: Key overlap ratio (\textbf{art.kor}) \cite{Bresin2000Articulation545}, measured at each note transition; overlap time (or gap time if \textit{staccato}) divided by the IOI between the two notes.    
    \item Dynamics: Comparing performed velocity and score marking (\f, \p, etc.), and measuring their agreement (\textbf{dyn.agr}) and consistency (\textbf{dyn.con}) as proposed by \citet{Kosta2018DynamicsScore}. We also propose the ramp correlation (\textbf{dyn.r\_cor}) for changing markings (hairpins) since \citet{Kosta2018DynamicsScore} only worked with constant markings. The ramp correlation computes the amount of agreement between the performed velocities with respect to their \textit{cresc.} or \textit{decresc.} ramp, if the markings exist. 
    \item Pedals: We measure the sustain pedal value at the note onset (\textbf{ped.onval}). Actually, sustain pedal change is a continuous stream of values and changes in pedal position often happen between note attacks; however, sampling at the note onsets simplifies the computation and allows for a consistent assessment across models. 
\end{itemize}

\subsubsection{Metrics}
For each expression dimension, we measure three metrics between the generated performance and the ground truth space:

\textbf{Standard deviation multiple}: This metric computes the deviation of an attribute of the rendered output from the mean of multiple human performances on a beat-level basis. Different from absolute deviation, this measure incorporates the flexibility of interpretations: when human interpretations already contain large differences, a larger discrepancy can be tolerated. But if human players tend to agree on the interpretation, we would expect the rendered output to fit more closely to ground truth values. Additionally, we retain the sign (direction) of deviation, so that negative values indicate deviations in the direction of slower tempo or softer dynamics, for example. 

\textbf{KL divergence} takes all the note-level or beat-level attributes, and compares their divergence with the ground truth attributes as an overall distribution. (Note that the ground truth attribute distributions are aggregated from multiple interpretations.) The KL divergence is calculated by estimating the two distributions using Monte Carlo sampling ($N=300$) and computing the relative entropy between them.

\textbf{Pearson's correlation} is measured between the feature sequences of generated and ground truth performances. In contrast to the previous two metrics, this metric captures the time-varying similarity between the performance attributes. 

\subsubsection{Results}

In Table~\ref{tab:quant_expression}, we compare our model with the samples from two existing performance rendering systems, BasisMixer \cite{CancinoChacon2018ComputationalModels} (BM) \footnote{We applied the Basis Mixer model with LSTM architecture, trained on the ASAP corpus.}, and VirtuosoNet \cite{Jeong2019VirtuosoNetPerformance} (VN) \footnote{The applied model is the \texttt{isgn} with default tempo and composer setting agreed by the author.}. Results are rendered on the same testing set as ours and shown as mean and standard deviation. 

Overall, DExter shows commendable results particularly in correlation across almost all performance dimensions, especially in capturing the global curve of tempo and velocity. This latter effect (learning the overall musical shape) could be attributed to the diffusion model predicting the time-varying codec in one pass, in contrast to autoregressive approaches. However, it is evident that DExter has room for improvement in terms of deviation and divergence: DExter's outputs demonstrate more volatile changes of parameters that are less smooth than other renderings. 

Meanwhile, each model exhibits distinct strengths across various performance dimensions. BM's outputs have articulation that is closer to human ground truths, and this can be attributed to BM being more conservative in its use of expressive devices, using smaller deviations from mechanical reproduction of the score. VN also excels in modelling the dynamics in agreement with the score markings. It is also notable that both models with sustain pedal prediction did a poor job in mimicking human pedaling techniques, with almost no time-wise correlation and on average one standard deviation away from the gound truth. Another area where all models struggle is the asynchrony time (asy.delta: $\mathtt{\sim}$4 deviations away from human benchmark), highlighting the need of refining the micro-timing aspect in performance rendering models.

\begin{figure}
    \centering
    \includegraphics[width=0.8\linewidth]{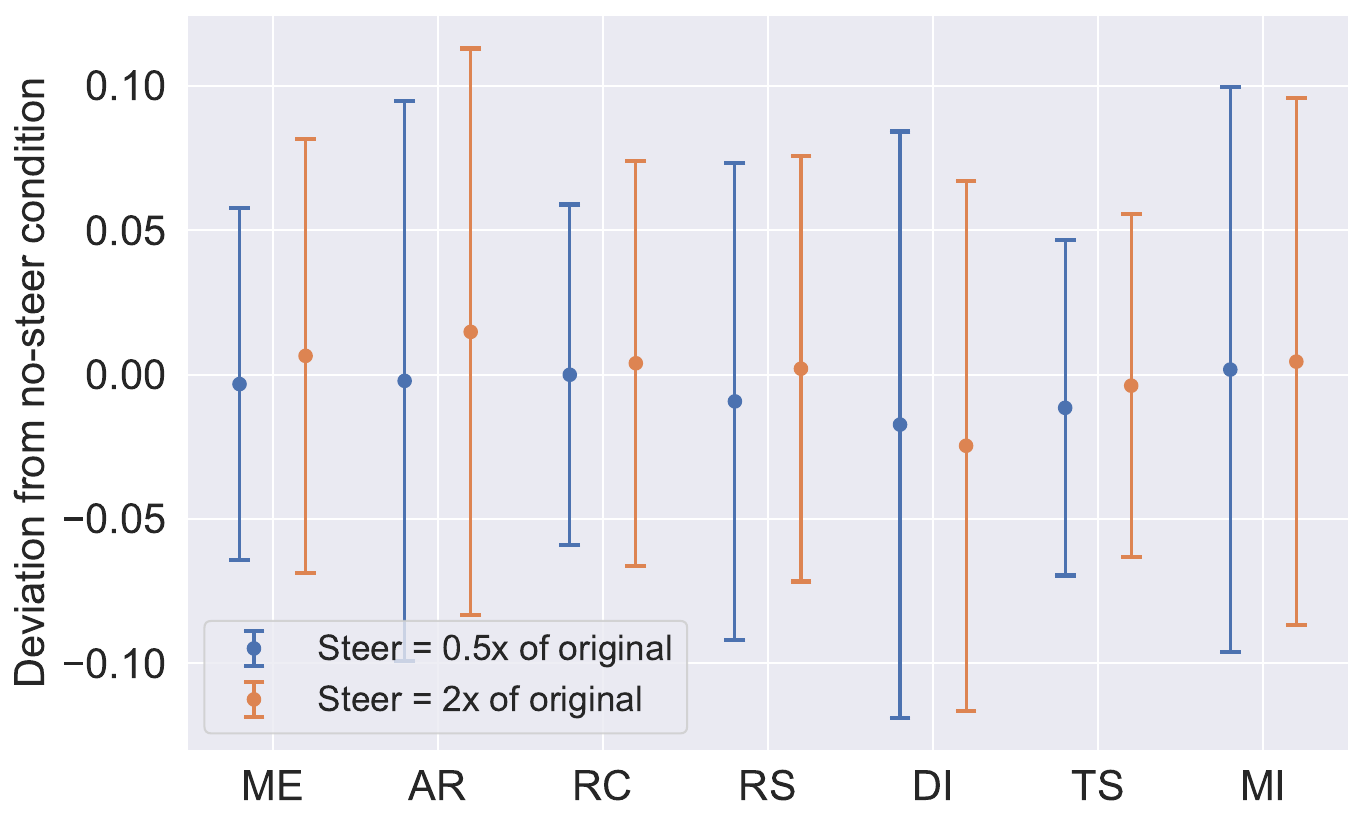}
    \caption{Steering the expressive characteristics of generated performances by using mid-level features (\texttt{c\_codec}, see Sec.\ \ref{sec:codecs}) as conditions. For each piece, a performance is generated with the \texttt{c\_codec} derived from an actual performance of the piece, and two further performances are generated with one of the mid-level features doubled (2$\times$) or halved (0.5$\times$). The average difference between the halved and unmodified, and between the doubled and unmodified conditions are plotted here.}
    \label{fig:ml_condition}
\end{figure}

\subsection{Expressive Steering with Perceptual Features}
\label{subsec:gen_target}

Our framework of conditioned performance generation allows us to explore conditioning the performance generation on additional features. As described in Sec.~\ref{sec:codecs}, we use mid-level perceptual features (encoded as \texttt{c\_codec}) as steering inputs to guide the expressive character of the generated performance. 

To gauge DExter's sensitivity to changes in these features we use the perceptual feature recognition model of \cite{Chowdhury2019} as a proxy for human perception. However, as that model had originally been trained on audio input, we wanted to eliminate the effect of acoustic artifacts introduced by rendering MIDI to audio, we decided to fit a MIDI-to-perceptual-features model to serve as the proxy instead. Details on this proxy model are given in the Appendix.

Steering performance generation is done by manipulating individual dimensions of the perceptual features, aiming to induce measurable corresponding effects in the resulting outputs. For each test sample and across all seven perceptual attributes, we first generate performances using the unmodified target
$c_c$. These target perceptual feature values could be randomly initialized in practice, or derived from an actual performance. We take the feature values derived from actual performances and modify the values to steer the generation in particular expressive directions, thus generating ``alternate" performances of the original performance. We either halve one feature, $\frac{1}{2}c_c$, or double one feature, $2c_c$, at a time.

Fig.~\ref{fig:ml_condition} displays the proxy model's predictions on the generated outputs. We observe that for the first four features \textit{melodiousness, articulation, rhythm complexity, rhythm stability}, the adjustments applied to the input conditions manifest as anticipated directional changes (the $c_{c, \mathit{double}}$ group leads the $c_{c, \mathit{half}}$ group 12.2\% in terms of their absolute value), providing evidence of the model's responsive behavior to the controlled feature alterations. 

The other three dimensions -- notably, \textit{dissonance} --
exhibit less consistent patterns in alignment with the input modifications. That seems reasonable, as harmonic and tonality-related properties are more a function of a piece itself, rather than any specific interpretation of it.

\subsection{Transferring from a Source Performance}
\label{subsec:gen_transfer}

As suggested by \citet{Liu2023AudioLDMModels} and \citet{Zhang2022Inversion-BasedModels}, style transfer can be achieved in a diffusion framework by using, as a  starting point for generation, a shallowly noised version of the source information. Given the large amount of music overlap in our datasets, we can test this by forming data pairs that consist of two interpretations of the same piece, to be used as the source and target \texttt{p\_codec} in this experiment. 

Given a source performance codec $x_\mathit{src}$, we calculate its noisy version $x_{t_0}$ with a predefined time step $t_0 \leq T$ according to the forward process shown in Equation~\ref{eq:q_sampling}. By using $x_{t_0}$ as the starting point for the reverse process of a pretrained model, we enable the manipulation of performance $x_\mathit{src}$ with target mid-level condition and shared score condition $c_\mathit{(s, tgt)}$ in a shallow reverse process $p_{\theta}(\hat x_{0:t_0} | x_{t_0}, c_\mathit{(s, tgt)})$, as illustrated in Fig.~\ref{fig:diffusion_diagram} (top, right). With the transfer experiments, we attempt to understand the following two questions:

\textit{1. Does transferring help with the final generation quality compared with rendering from scratch?}

\begin{table}
\centering
\begin{tblr}{
  hline{1-2,6} = {-}{},
}
$t$            & \textit{dev}: tem  ($\downarrow$)  & \textit{dev}: vel  ($\downarrow$)  & \textit{cor}: tem  ($\uparrow$) & \textit{cor}: vel ($\uparrow$) \\
$T$            &  0.72 $\pm$2.65  &   1.48$\pm$ 2.13   &   0.19$\pm$0.17  &  0.27$\pm$0.23      \\
$\frac{3T}{4}$ &  \textbf{0.68$\pm$2.55}   &   1.40$\pm$2.16   &   \textbf{0.19$\pm$0.16}   &   \textbf{0.28$\pm$0.21} \\
$\frac{T}{2}$ &  0.74$\pm$2.49   &   \textbf{1.33$\pm$2.10}   &   0.15$\pm$0.17   &   0.21$\pm$0.22      \\
$\frac{T}{4}$ &  0.87$\pm$2.69   &   1.50$\pm$2.11   &   0.11$\pm$0.16   &  0.18$\pm$0.21        
\end{tblr}
\caption{Deviation and correlation of test set relative to ground truth space (same analysis as in Section~\ref{subsec:quant_analysis}) of tempo and velocity curves.\label{tab:transfer_quality}}
\end{table}

In the transfer experiment, we combine pairs of ground truth performances of the same piece segment $x_\mathit{src}, x_\mathit{tgt}$ where $s_\mathit{src} = s_\mathit{tgt}$. The same testing set as the previous sections is used, and the source performance is randomly taken from the ground truth.  We experimented with different transfer steps of $t_0 \in \{T, \frac{3T}{4}, \frac{T}{2}, \frac{T}{4}\}$, and report the global metrics of tempo curve and velocity curve with their deviation and correlation. 

What we observed in Table~\ref{tab:transfer_quality} is that transferring from a source performance slightly helps with initialization. Specifically, employing a denoiser for three-quarters of the diffusion steps -- ideally preserving around one-quarter of the source's characteristics -- yielded the highest quality outcomes. 
However, transfer quality does not steadily improve with the retained information from source: the $\frac{T}{4}$-step transfer results in ambiguous outputs that do not align well with the given score.

\textit{2. Does a transferred rendering sound `closer' to the source or the target?}  

\begin{figure}
    \centering
    \includegraphics[width=0.8\linewidth]{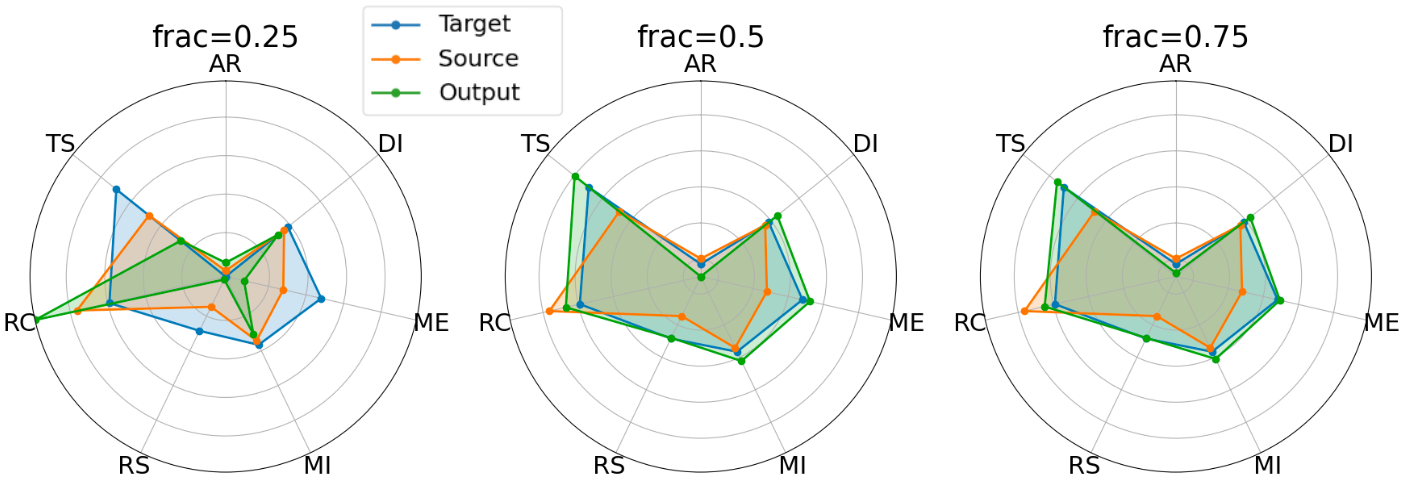}
    \caption{Seven dimensions of perceptual features (AR: articuation; RC: rhythm complexity; RS: rhythm stability; TS: tonal stability; DI: dissonance; MI: Minorness; ME: Melodiousness) predicted using the proxy for output, source, and target, averaged across the testing set. 
    The three plots correspond to transfer steps of $0.25T, 0.5T, 0.75T$.
   }
    \label{fig:transfer_radar}
\end{figure}

Similar to Sec.~\ref{subsec:gen_target}, we wish to measure the transfer proximity using the predicted perceptual features. The radar plots in Fig.~\ref{fig:transfer_radar} show the seven perceptual feature dimensions predicted by the proxy, illustrating the perceptual distance between source, target, and generated performance for three different transfer gradations, $\frac{T}{4}, \frac{T}{2}, \frac{3T}{4}$. At $\frac{T}{4}$ steps, the predicted performance deviates from both source and target, which fits our previous observation that insufficient denoising steps result in ambiguous outputs. As the transfer step increases, there is a discernible shift in the predicted output towards the target profile across most perceptual dimensions.

\subsection{Effect of Varying Conditioning Weights}
\label{subsec:conditioning}

In this experiment, we look at the effect of the conditioning weight $w$ on the generated results. As described in Sec.~\ref{sec:conditioning} and Eq.~\ref{eq:weight}, the scale of classifier-free guidance $w$ is the ratio that combines the prediction with and without (masked by 0) the $c_s$ and $c_c$ conditioning. While the conditional and unconditional models are jointly trained in the training phase, the weighting parameter $w$ is only introduced in the sampling phase and the optimal $w$ is not trivial to find. In Tab.~\ref{tab:w_quality}, conditioning weights $w=0.5, 1.2, 2, 3$ are compared, while other settings are the same as in Section~\ref{subsec:quant_analysis}. 
Experimental results are best for $w=1.2$. 
Interestingly, with greater scale of classifier guidance, the generative results exhibit larger fluctuations in expressive parameters and less stability.

\begin{table}
\centering
\begin{tblr}{
  hline{1-2,6} = {-}{},
}
$w$            & \textit{dev}: tem ($\downarrow$)  & \textit{dev}: vel($\rightarrow$0)  & \textit{cor}: tem($\uparrow$)  & \textit{cor}: vel ($\uparrow$) \\
$0.5$            &  1.11 $\pm$2.46  &   -2.47$\pm$ 1.10   &   0.11$\pm$0.16  &  0.02$\pm$0.22      \\
$1.2$ &  \textbf{0.72$\pm$2.65}   &   \textbf{1.48$\pm$2.12}   &   \textbf{0.19$\pm$0.16}   &   \textbf{0.28$\pm$0.21}      \\
$2$ &  1.33$\pm$2.37   &    3.02$\pm$1.53   &   0.04$\pm$0.15  &   0.13$\pm$0.24     \\
$3$ &  1.86$\pm$1.81   &   4.63$\pm$1.15   &   0.04$\pm$0.14   &  0.10$\pm$0.23             
\end{tblr}
\caption{Deviation and correlation of test set relative to ground truth space (same analysis as in Section~\ref{subsec:quant_analysis}) of tempo and velocity curves.\label{tab:w_quality}}
\end{table}

\subsection{Qualitative study}

We evaluated the naturalness and expressiveness of the rendered performances through a listening test. For samples from eight selected pieces, we compared the following: 1) two human performances, with relatively distant interpretations; 2) renderings made with Basis Mixer and VirtuosoNet, as described in Section \ref{subsec:quant_analysis}; and 3) a rendering from the proposed model DExter. The performances (including the ground truths) were rendered to audio using a Yamaha Disklavier, which produced similar pedal/articulation-related artifacts both in the human and the machine performances. 82 participants listened to the performances and evaluated them on a 100-point Likert scale, rating the overall naturalness and expression of the output as one score. The performances used for the test can be found on the demo page. 

\begin{figure}
    \centering
    \includegraphics[width=0.9\linewidth]{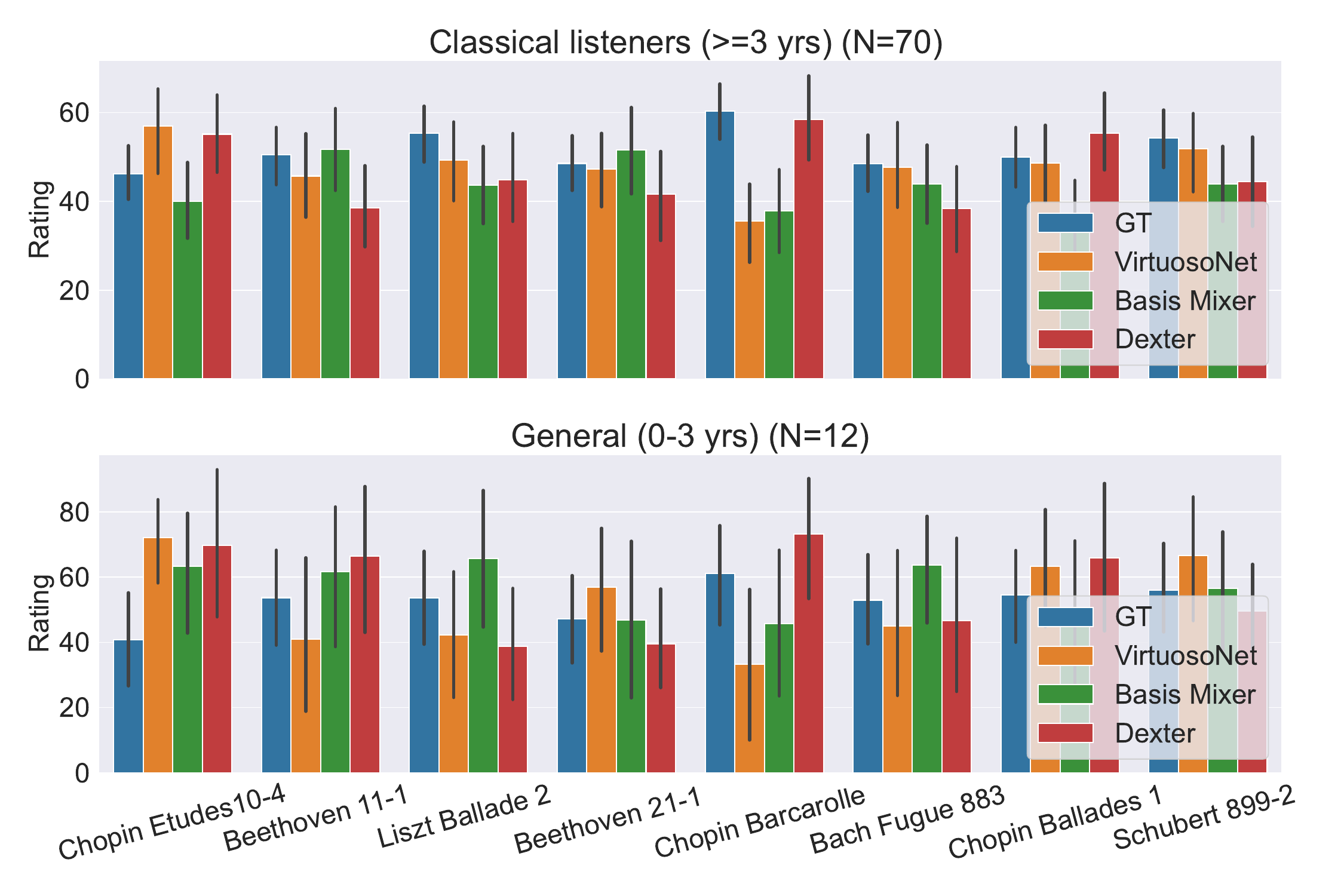}
    \caption{Mean ratings for the eight pieces evaluated in the listening test, for each model and human (Ground Truth, GT) performances.
    }
    \label{fig:listening_result}
\end{figure}

The results of the listening test, shown in Fig.~\ref{fig:listening_result}, give a nuanced view of the performance rendering capabilities of the models in comparison to the ground truth. In terms of the mean rating, DExter demonstrates better performances of the pieces of Chopin, even sometimes comparable to the ground truth (\textit{Barcarolle}). However, it is outperformed by Basis Mixer or VirtuosoNet in the case of older compositions (\textit{Bach Fugue} and \textit{Beethoven Sonatas}). Overall, in terms of the mean rating scores, there is still a gap between the generative outputs and GT (51.81), while DExter (48.54) slightly outperforms VirtuosoNet (48.31) and Basis Mixer (46.33). It is also surprising to observe that GT does not always secure the highest ratings. In the case of the Chopin Etude, at least, it might be explained by the fact that the human pianists suffer from physical limitations in technically demanding passages, while the generative models do not.

Besides the numerical ratings, we also asked three musically trained participants for explicit feedback on their ratings. In addition to some positive comments, we also received quite specific and useful negative feedback, such as (specifically referring to polyphonic music such as Bach) \textit{no clear voicing among the lines} and \textit{poor balance between hands}. Given currently dominating `flattened' representations such as our \texttt{p\_codec} or tokens in Transformer models, learning the vertical structure of music remains a challenge to rendering models \cite{Zhang2023SymbolicEvaluation}.   

%


\section{Conclusion and future work}


In this paper, we introduce a novel diffusion-based model, DExter, for learning and controlling performance expression in solo piano music. Besides rendering performances at a comparable level of quality to existing models in quantitative measurement of expressive characteristics, DExter is also capable of style transfer between interpretations and conditioning the rendered expression with perceptual variables. 

As a future direction, we would like to improve on the inference speed of DExter (currently 40 seconds inference time for a 95 second piece on single RTX 6000 GPU). To accelerate sampling of 1k step iterative process we would explore techniques like DDIM or learning in a latent space. On the other hand, we would also like to explore other conditioning inputs like text that could allow for more explicit controls over the rendered outputs.

\section{Acknowledgment}
This work is supported by the UKRI Centre for Doctoral Training in Artificial Intelligence and Music, funded by UK Research and Innovation [grant number EP/S022694/1], also by the European Research Council (ERC) under the EU's Horizon 2020 research and innovation programme, grant agreement No.~101019375 (\textit{Whither Music?}). We also like to thank Daesam Jeong for contributing VirtuosoNet and for comparative discussion, as well as Maximillian Hofmann for the Basis Mixer trained with LSTM.

\section[\appendixname~\thesection]{Appendix}
\subsection[\appendixname~\thesubsection]{MIDI to perceptual features model}
The MIDI to perceptual features model is used as proxy for human mid-level perception in the experiments described in Sec.~\ref{subsec:gen_target} and Sec.~\ref{subsec:gen_transfer}. It takes in a rendered MIDI and outputs a 7-dimensional perceptual features, for each window of 15 seconds. The specifications are as follows:

\begin{itemize}
    \item Data: The data used to train this oracle is ASAP performance MIDI, along with the audio perceptual features computed predicted from ASAP performance audio by the mid-level feature recognition model of \cite{Chowdhury2021OnFeatures}, for 15 seconds windows. 
    \item Representation: Each 15 seconds MIDI window is transformed into a piano-roll matrix of dimension 800 * 131 (128 pitches + 3 pedal channels), with MIDI velocity as matrix value. 
    \item Architecture: The network consists of two residual blocks, each containing two convolution layers. A final projection layers is attached at the end to output the 7-dimension perceptual features.
    \item Training: Adam optimizer with learning rate $1e-3$. After 20 epochs the training converges with validation loss 0.038. 
\end{itemize}

\begin{adjustwidth}{-\extralength}{0cm}

\reftitle{References}

\bibliography{ref}

\PublishersNote{}
\end{adjustwidth}
\end{document}